\documentclass[%
 reprint,
 groupedaddress,
 amsmath,amssymb,
 aps,
]{revtex4-1}

\usepackage[utf8]{inputenc}

\usepackage[normalem]{ulem}
\usepackage{float}
\usepackage{color}
\usepackage[dvipsnames]{xcolor}
\usepackage{graphicx}
\usepackage{dcolumn}
\usepackage{bm}

\newcommand{\rcoor}{{r}}
\newcommand{\rev}[1]{{{#1}}}

\begin{document}

\preprint{APS/123-QED}

\title{{Observation of oscillatory radial electric field relaxation in a helical {plasma}}}

\author{J. A. Alonso}
\author{{E. S\'anchez}}
\author{{I. Calvo}}
\author{{J. L. Velasco}}
\author{{T. Estrada}}
\author{{K. J. McCarthy}}
\author{{P. Monreal}}
\author{{the TJ-II Team}}
\affiliation{Laboratorio Nacional de Fusi\'on, CIEMAT, 28040, {Madrid,  Spain}}
\author{S. Perfilov}%
\author{L. G. Eliseev}%
\author{A. V. Melnikov}
\affiliation{%
 National Research Center, Kurchatov Institute, 123182, {Moscow, Russia}}%
\author{A. Chmyga}%
\author{L. I. Krupnik}%
\author{A. I. Zhezhera}
\affiliation{%
  Institute of Plasma Physics, NSC KIPT, 310108, Kharkov, Ukraine}%
 \author{F. I. Parra}
\affiliation{Rudolf Peierls Centre for Theoretical Physics, {University of Oxford,} Oxford,  OX1 3NP, {United Kingdom}}
 \author{R. Kleiber}
\affiliation{Max-Planck Insitut f{ü}r Plasmaphysik, Greifswald, 17491, Germany}


\date{\today}

\begin{abstract}
Measurements of the relaxation of a zonal electrostatic potential perturbation in a non-axisymmetric magnetically confined plasma are presented. A sudden perturbation of the plasma equilibrium is induced by the injection of {a cryogenic hydrogen pellet} in the TJ-II stellarator, which is observed to be followed by a damped oscillation in the electrostatic potential. The waveform of the relaxation is consistent with theoretical calculations of zonal potential relaxation in a non-axisymmetric magnetic geometry. 
{The turbulent transport properties of a magnetic confinement configuration are expected to depend on the features of the collisionless damping of zonal flows, of which the present {letter} is the first direct observation.}

\end{abstract}

\pacs{Valid PACS appear here}
\maketitle

\section{Introduction}
One of the main challenges of nuclear fusion by magnetic confinement is to confine the {plasma} particles {and energy} long enough for fusion reactions to occur in sufficient number in the central, hottest and {densest} part of the confining toroidal magnetic structure. That structure consists of a strong magnetic field that is everywhere tangent to a family of nested tori, the so-called {flux surfaces}. %
The achievable central plasma pressure depends upon the characteristics of the energy transport across the flux surfaces. In axisymmetric devices like tokamaks, ion temperature gradient (ITG) driven {microturbulence} is a common cause of cross-field diffusion. The linear characteristics of ITG instability (namely its large growth {rates} and small radial wavelengths) make it very deleterious for confinement and would pose a serious limit to the performance and economic viability of a fusion reactor. Fortunately enough, the non-linear evolution of the instability produces zonal electrostatic  {perturbations (i.e. perturbations in the electrostatic potential that are constant on each flux surface)} which, while not causing any cross-field transport by themselves, {reduce} the radial correlation of turbulent eddies by 
$E\times B$ shearing \cite{Biglari90}.

The non-linear generation of zonal flows by drift-wave type turbulence is known to be robust {\cite{Hasegawa87}}, but the question of whether those flows were strongly damped in toroidal geometry and therefore unable to tame turbulent transport troubled the fusion community for some time \cite{GlanzScience1996}. {In their seminal work \cite{RosenbluthPRL1998}, Rosenbluh and Hinton  showed that long-wavelength zonal flows} were not completely suppressed by collisionless Landau-type damping in axis-symmetric toroidal devices like tokamaks. The initial value problem of the collisionless relaxation of a zonal electrostatic potential {is known} as the Rosehbluth-Hinton problem and {has become} a standard test for gyrokinetic codes \cite{Garbet2010}.

{In recent years, the interest in the Rosenbluth-Hinton problem has been fostered by its extension to stellarator geometry \cite{SugamaPoP2006, HelanderPPCF2011, MonrealPPCF2016}. Stellarators are a type of toroidal magnetic confinement device whose magnetic field is created solely by external coils. This, in turn, requires to break the {axisymmetry} of the flux surfaces to achieve magnetohydrodynamic equilibrium and stability of the confined plasmas. Unlike in tokamaks, long-wavelength zonal flows are strongly damped in stellarators even in the absence of collisions. Furthermore, their {collisionless} time evolution is different to that in tokamaks in that they exhibit a low frequency oscillation --different from the geodesic acoustic mode also observed in {tokamaks-- that} accompanies the relaxation to the {final} \emph{residual} level~\cite{MishchenkoPoP2008}.  This oscillation constitutes a distinct feature of the collisionless electric field relaxation in three-dimensional devices susceptible 
to experimental 
detection and, at the same time, it can be more relevant for the moderation of turbulent transport than the behaviour of the zonal perturbation at infinite time.}

{It has been argued that the properties of the linear relaxation can influence the non-linear saturated turbulent heat transport \cite{WatanabePRL2008, XanthopoulosPRL2011}, which would support the use of those linear properties as a computationally inexpensive figure of merit in search of turbulence-optimized magnetic configurations  for future nuclear fusion reactors. In this letter we have taken a step further by presenting direct observations of the oscillatory relaxation of a zonal electrostatic potential perturbation in a stellarator plasma, that are shown to be in qualitative agreement with {gyrokinetic} simulations. To our knowledge, this is the first experimental confirmation of the relevance of the linearised gyrokinetic equation to describe zonal flow relaxation in toroidal magnetized plasmas.}

\section{Theory}
The {description} of the {collisionless} relaxation of a zonal electrostatic potential perturbation makes use of {gyrokinetics, the kinetic theory for strongly magnetized particles that is obtained by systematically averaging the fast quasiperiodic gyromotion of the particles around a magnetic field line to obtain a kinetic equation for the particles' gyrocentres}.

In strongly magnetized fusion plasmas, the particles of any species $s$ are confined for longer than a collision time and, conseguently, their distribution function is close to a Maxwellian $f_{Ms}$ with temperature $T_s$ and density $n_s$  constant on flux surfaces.  
In addition, {the magnetization parameter} $\rho_{ts} / L \ll 1$ is the basis of the scale separation assumption in local gyrokinetic theory, based on which we can write the distribution function and the electrostatic potential in eikonal form. {Here $\rho_{ts} = v_{ts} / \omega_s$ is the Larmor radius of species $s$, where $v_{ts} = \sqrt{T_s/m_s}$ is the thermal speed, $m_s$ is the mass, and $\omega_s = Z_s e B / m_s$ is the gyrofrequency, $e$ is the proton charge, $Z_s e$ is the charge of species $s$ and $B$ is the magnitude of the magnetic field $\mathbf{B}$}. Finally, since the gyrokinetic equation {for purely zonal perturbations is linear, we} 
can study the 
evolution of a single mode. That is, using an effective radius $\rcoor(\mathbf{x})$ as the flux surface label, we can take $\varphi(\mathbf{x},t) =
\varphi_k(\rcoor(\mathbf{x}),t) \exp(\mathrm{i}k_\rcoor\rcoor(\mathbf{x})),$ where $L^{-1}\ll k_r \lesssim \rho_{ts}^{-1}
$ and $\varphi_k$ varies on the scale $L$. 
Analogously, denoting by $F_{1s}$ the deviation of the distribution function from $f_{Ms}$, we write   $F_{1s}(\mathbf{x},v,\lambda,\sigma,t)=
  f_s(\rcoor(\mathbf{x}),\theta(\mathbf{x}),\alpha(\mathbf{x}),
v,\lambda,\sigma,t) \,\exp({\mathrm{i}k_\rcoor\rcoor(\mathbf{x})})$. {We are employing} the magnitude of the velocity $v$, the pitch-angle coordinate $\lambda = v_\perp^2 /(v^2 B)$ and the sign of the parallel velocity $\sigma$ as coordinates in velocity space, where $v_\perp$ is the magnitude of the component of the velocity perpendicular to the magnetic field. Then, the {gyrokinetic equation for the above kind of perturbations} reads
\begin{equation}\label{eq:gyrokineticequation}
 \left(\partial_t + v_\parallel\, \hat {\bf b} \cdot \nabla 
  + \mathrm{i}k_\rcoor {v_{r,s}}\right) h_s 
 = \frac{Z_s e}{T_s} \partial_t \varphi_k J_{0s} f_{Ms},
\end{equation}
where $v_\parallel = \sigma v\sqrt{1- \lambda B}$, $\hat{\bf b} = \mathbf{B}/B$ and $h_s:= f_s + (Z_s e/T_s) \varphi_k J_{0s} f_{Ms}$.  Here, $J_0$ is the zeroth-order Bessel function of the first kind and ${v_{r,s}} = \mathbf{v}_{Ms}\cdot\nabla r$, with $\mathbf{v}_{Ms}$ being  the magnetic drift of species $s$. We have eased the notation by writing $J_{0s} \equiv J_0(k_\rcoor |\nabla \rcoor| \rho_s)$, with $\rho_s = v_\perp / \omega_s$. Both $f_s$ and $h_s$ vary on the scale $L$. 

{The solution} of (\ref{eq:gyrokineticequation}) at long times coupled to the quasineutrality equation
{\begin{equation}\label{eq:qnequation}
 \sum_s \frac{Z_s^2 e}{T_s}n_s\, {\varphi}_k = 
  \left\langle 
   \sum_s Z_s \int J_{0s} {h}_s
    \mathrm{d}^3 v  
  \right\rangle_\rcoor
\end{equation}
 has been  derived in detail in \cite{
MonrealPPCF2016}. Here, $\langle \ \cdot \ \rangle_r$ denotes flux surface average. Taking the limit of small radial drift frequency in the expression for the long time evolution of the electrostatic potential, one finds that $\varphi_k$ oscillates with frequency
\begin{equation}\label{eq:omegakinspecies}
\Omega = \sqrt{\frac{A_2}{A_1 + A_0}},
\end{equation}
where $A_0 = \sum_s n_s \frac{Z_s^2 }{T_s}
\left\langle |\nabla\rcoor|^2\rho_{ts}^2\right\rangle_\rcoor$, $A_1 = \sum_s \frac{Z_s^2 }{T_s}\left\{\delta_s^2 
- \overline{\delta_s}^2\right\}_s$ and $A_2 = \sum_s \frac{Z_s^2 }{T_s}\left\{\overline{{v_{r,s}}}^2\right\}_s$, and where $\left\{Q\right\}_s  := 
 \left\langle
 \int Q f_{Ms} \mathrm{d}^3v
  \right\rangle_\rcoor$, $Q$ being a function on phase-space. An overline stands for the orbit average operation. We have split the radial magnetic drift into its orbit average and a component that varies along the orbit, ${v_{r,s}} = \overline{{v_{r,s}}} + v_\parallel \hat {\bf b} \cdot \nabla  \delta_s$. Observe that $\delta_s$ can be interpreted as the radial excursion of the particle at each point of its orbit. The oscillatory behavior of zonal flow relaxation in stellarators was originally pointed out in \cite{MishchenkoPoP2008} and explained in more detail in  \cite{HelanderPPCF2011}. In these references, (\ref{eq:omegakinspecies}) was derived and its quantitative accuracy as solution of (\ref{eq:gyrokineticequation}) and (\ref{eq:qnequation}) is proven in \cite{MonrealInPrep}. In \cite{HelanderPPCF2011}, an approximate expression for the collisionless damping rate of the oscillations is provided as well.}
  
{From the explicit expression of the frequency (\ref{eq:omegakinspecies})} it is obvious why the oscillation does not exist in tokamaks. Namely, because $\overline{{v_{r,s}}}$ vanishes identically in axisymmetric magnetic fields. Furthermore, in the derivation {of this expression} {one learns} that the oscillation is expected to be observed if the scale of the zonal perturbation $k_r^{-1}$ is sufficiently large for the {small radial drift frequency condition} $|k_\rcoor\overline{{v_{r,s}}}|/\Omega \ll 1$ to be satisfied, and sufficiently small for (\ref{eq:gyrokineticequation}) to be applicable.
{It is also {important to note} that the oscillation with frequency (\ref{eq:omegakinspecies}) is a collisionless phenomenon, and that introducing collisions in the system does not modify it  unless the collision frequency is much larger than $\Omega$. However, the damping rate of the zonal flow is very sensitive to the collision frequency.}

\section{Experimental set-up and observations}
\begin{figure}
\centering
\includegraphics{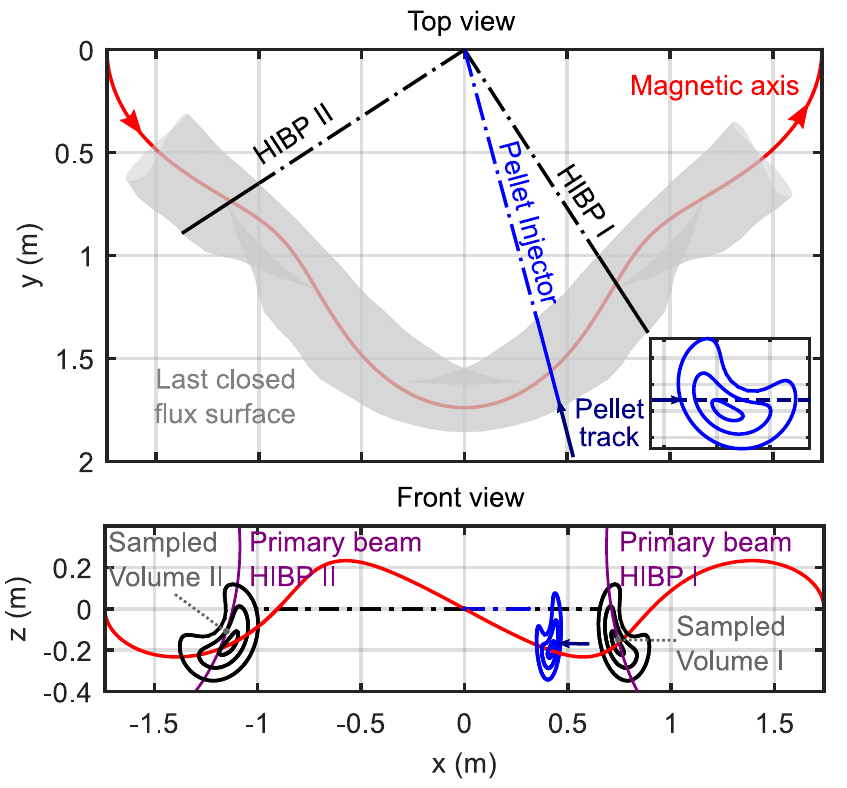}
\caption{Schematic view of TJ-II magnetic geometry and  experimental set-up. The positions of the pellet injection line and the two HIBP systems in equal cross-sections of two consecutive periods are shown. The upper plot shows a top view of one half of toroidal magnetic structure of the TJ-II stellarator. The magnetic axis is shown in red and a section of the last closed flux surface is shown in gray. The inset shows the cross section of several flux surfaces at the poloidal location of the pellet injector. These and the cross sections at the location of the HIBPs, are shown in a front view in the lower plot.}
\label{fig:Geometry}
\end{figure}

Plasmas studied in this {letter} were produced in the TJ-II stellarator (4 periods, average field on axis 0.96 T, edge rotational transform 1.65) and heated by neutral beam injection. The discharges used here have line-averaged electron densities between {$1-2.5\times 10^{19}$} particles/{m}$^3$ and central electron and ion temperatures of 250 and 100 {eV, respectively}. A cryogenic hydrogen pellet is injected into an otherwise {effectively} stationary phase of each discharge, causing a rapid change of the plasma parameters. The transient dynamics of the electric field that follows the pellet ablation is studied by means of two heavy ion beam probe (HIBP) systems \cite{MelnikovFST2007, MelnikovNF2011}. A schematic view of the setup showing the relative location of the pellet injection line and the two probing systems is presented in figure \ref{fig:Geometry}.
The HIBP diagnostics were set to measure plasma potential at fixed points of the plasma cross section (termed `sampled volume' in figure \ref{fig:Geometry}) with spatial resolution of $\lesssim 1.5$ cm  and 2 MHz sampling rate. 

\begin{figure}
\centering
\includegraphics{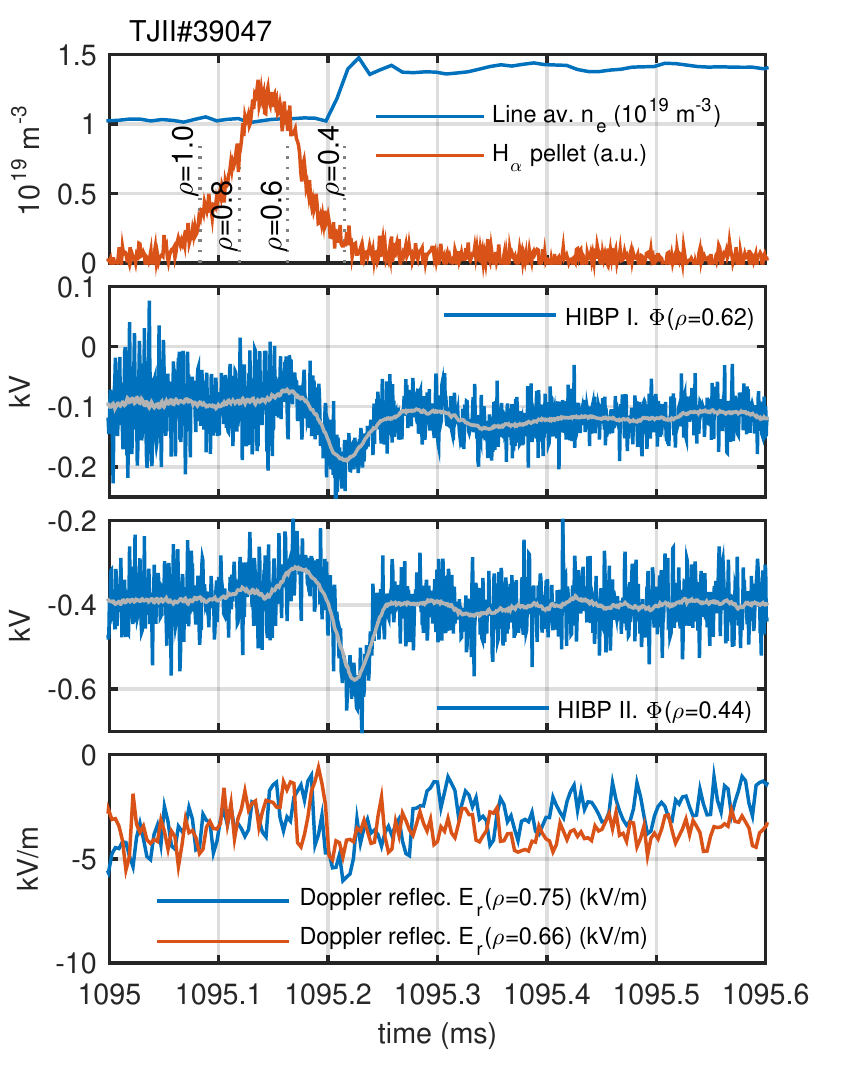}
\caption{Evolution of plasma parameters following a pellet injection into the TJ-II stellarator. Top: line averaged electron density and $H_\alpha$ pellet monitor. Center: plasma electrostatic potential at two radial locations from the HIBP diagnostic. Bottom: radial electric field in two peripheral radial locations from Doppler reflectometry \cite{HappelRSI2009}.}
\label{fig:PelletInjection}
\end{figure}

Figure \ref{fig:PelletInjection} shows several diagnostic signals in a 600 microseconds time window during a pellet injection. 
\rev{The ablation of the pellet, from its entry into the plasma until its complete ablation, is registered with silicon diodes that measure the line radiation (at 656.3 nm) emitted from the neutral cloud that surrounds the pellet \cite{MccarthyECPD2015}.}

Pellets are injected with velocities of $\sim 1000\,${m/s} and contain between $5\times 10^{18}$ and $10^{19}$ particles when condensed in the injector. Whereas the actual size of the pellets at first contact with the plasma edge varies, the estimated fuelling efficiency is about \rev{{$50\%$.}} The injections cause a prompt $\sim10\%$ reduction of the electron and ion temperatures as the pellet travels into the central plasma, and a $\sim{40\%}$ increase of the line-averaged plasma density. The central temperatures return to their previous levels after $5\,$ms, while the evolution of the line-averaged density is significantly slower (not shown in figure~\ref{fig:PelletInjection}, see ~\cite{VelascoPPCF2016}).
The plasma electric field undergoes a rapid evolution as observed with Doppler reflectometry in the plasma edge and the HIBP I and II in more interior positions (Fig.\ref{fig:PelletInjection}).  The high time resolution and good signal-to-noise ratio of the beam probes allow to resolve a damped oscillation in the electrostatic potential transient response, qualitatively similar to the relaxation typically observed in numerical Rosenbluth-Hinton tests in three dimensional magnetic fields \cite{MishchenkoPoP2008, SanchezPPCF2013}. 

\section{Comparison with linear gyrokinetic simulations}
\begin{figure}
\centering
\includegraphics{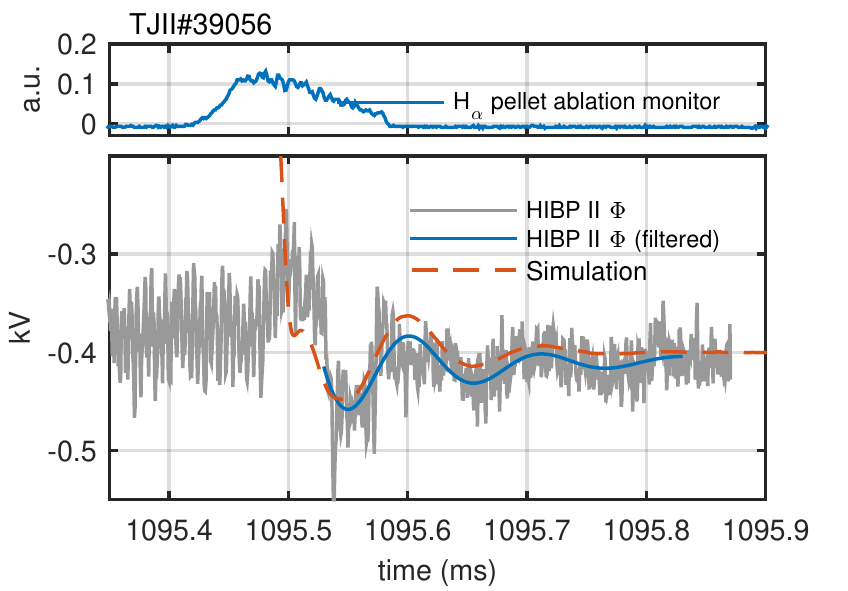}
\caption{{Time traces of the $H_\alpha$ ablation monitor in a pellet injection (top) and of the electrostatic potential in a Rosenbluth-Hinton gyrokinetic numerical simulation and in a pellet-induced transient in the TJ-II stellarator (bottom).}}
\label{fig:CompRHpellet}
\end{figure}

A direct comparison of the measured and simulated relaxation is shown in figure \ref{fig:CompRHpellet}, where the raw HIBP signal has been filtered to highlight the similarity of the low-frequency waveform with the result of the linear gyrokinetic simulation. The beginning of the simulation shows a quickly damped geodesic acoustic mode followed by the lower frequency damped oscillation as first {predicted in \cite{MishchenkoPoP2008}}.  

The simulations are carried out with the gyrokinetic Monte Carlo code EUTERPE \cite{JostPoP2001} in the full radial domain of the TJ-II standard magnetic configuration. The experimental density and temperature profiles together with the background radial electric field, obtained with the drift-kinetic equation solver DKES \cite{HirshmanDKES}, are used to define the equilibrium state. 
The run is initiated with a radially localized perturbation to the equilibrium ion probability distribution function with Maxwellian velocity dependence. The perturbed distribution function and electrostatic potential are then evolved according to the linearized gyrokinetic equations \cite{KleiberCPP2010}. The zonal component of the electrostatic potential at the flux surface of interest is monitored, which {corresponds} to the simulation 
values displayed in figure \ref{fig:CompRHpellet}. Both, collisionless and also collisional simulations have been carried out {confirming} that the oscillation frequency is basically determined by collisionless processes and not much affected by collisions, while the damping rate is largely dependent on the collisionality.

\begin{figure}
\centering
\includegraphics{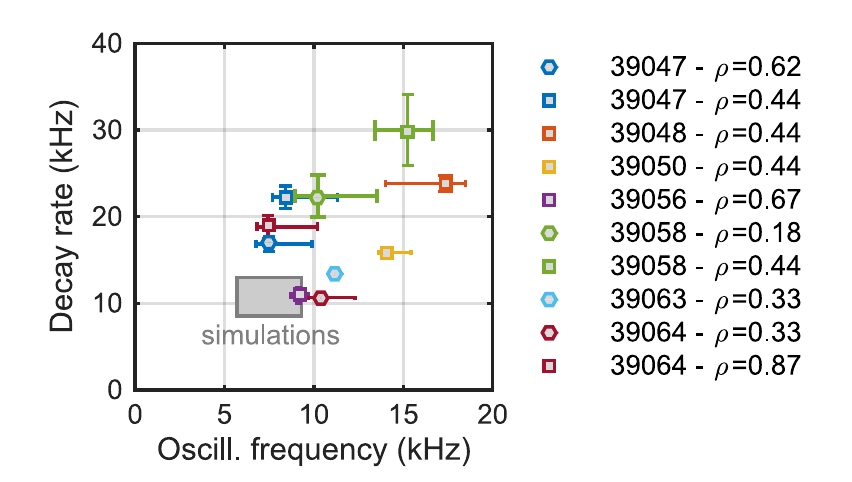}
\caption{Comparison of measured and simulated oscillation frequencies and  damping rate for several discharges and radial positions. Measurements obtained with the HIBP I  and II are labeled with circles and squares receptively. The gray rectangle encompasses the values obtained from the simulated zonal potential relaxation for several radii and five different experimental profile sets representative of the discharges.}
\label{fig:omegagamma}
\end{figure}

To systematize and quantify the comparison shown in figure \ref{fig:CompRHpellet}, values of the oscillation frequency and decay time are extracted from both the simulated and measured damped oscillation. This is done for the database of pellet injections in which a damped oscillation is visible. We note that not all pellet injections cause a measurable oscillation in the HIBP potentials, which is attributed to the imperfect reproducibility of the pellet and the different target plasma conditions.
The results are presented in figure \ref{fig:omegagamma}, with error bars corresponding to 95\% confidence level in the estimation of the frequency and damping rate. The rectangle {covers} the same confidence level for the parameters extracted from the decays simulated for several experimental discharges and the range of radial positions of experimental measurements. The agreement between the parameters of the simulated and measured relaxations is reasonable. In general, both the simulated damping rate and oscillation frequency show a shift towards smaller values compared to the measurement. Oscillation frequency displays a better agreement (within a factor of 2) and less dispersion than the damping rates. We note that the former, which is essentially determined by collisionless mechanisms, {is expected to be less affected by the uncertainties in} the local plasma parameters \rev{and plasma composition \cite{BraunPPCF2009}} as compared to the latter.

\section{Conclusions}
In conclusion, we have presented measurements of {zonal} electrostatic potential relaxation in pellet-induced fast transients which are consistent with the theoretical expectations and simulations of {gyrokinetic} theory.  These results are important in as much as the non-linear turbulence saturation level and heat transport coefficients in stellarators depend on the features of the collisionless damping of zonal electrostatic potential perturbations, of which the present {letter} is the first direct observation. The methodology and observations presented here can be reproduced in existing helical devices like the recently started {Wendelstein 7-X stellarator,} whose magnetic configuration and {low collisionality plasma conditions} {might} allow an {even clearer} observation of this phenomenon. 

\section*{Acknowledgments}
{The first author acknowledges A. Mishchenko for useful comments on the original manuscript. This work has been carried out within the framework of the EUROfusion Consortium and has received funding from the Euratom research and training programme 2014-2018 under grant agreement No 633053. The views and opinions expressed herein do not necessarily reflect those of the European Commission. This research was supported in part by grants ENE2012-30832 and ENE2013-48679, Ministerio de Econom\'ia y Competitividad, Spain.}The work of AVM was partly supported by the Competitiveness Programm of NRNU MEPhI. {The simulations were carried out within the Red Espa\~nola de Supercomputaci\'on infrastructure.}

\bibliography{bibliography}{}

\end{document}